\documentclass[twocolumn,conference]{IEEEtran}
\usepackage[T1]{fontenc}
\usepackage{amsmath}
\usepackage{amsthm}
\usepackage{amssymb}
\usepackage[unicode=true,
 bookmarks=true,bookmarksnumbered=true,bookmarksopen=true,bookmarksopenlevel=1,
 breaklinks=false,pdfborder={0 0 0},backref=false,colorlinks=false]
 {hyperref}
\hypersetup{pdftitle={Your Title},
 pdfauthor={Your Name},
 pdfpagelayout=OneColumn, pdfnewwindow=true, pdfstartview=XYZ, plainpages=false}

\makeatletter
\theoremstyle{plain}
\newtheorem{thm}{\protect\theoremname}
\theoremstyle{definition}
\newtheorem{defn}[thm]{\protect\definitionname}
\theoremstyle{plain}
\newtheorem{prop}[thm]{\protect\propositionname}
\theoremstyle{remark}
\newtheorem{rem}[thm]{\protect\remarkname}
\theoremstyle{plain}
\newtheorem{lem}[thm]{\protect\lemmaname}
\theoremstyle{definition}
\newtheorem{example}[thm]{\protect\examplename}
\theoremstyle{plain}
\newtheorem{cor}[thm]{\protect\corollaryname}

\usepackage[caption=false,font=footnotesize]{subfig}

\makeatother

\providecommand{\corollaryname}{Corollary}
\providecommand{\definitionname}{Definition}
\providecommand{\examplename}{Example}
\providecommand{\lemmaname}{Lemma}
\providecommand{\propositionname}{Proposition}
\providecommand{\remarkname}{Remark}
\providecommand{\theoremname}{Theorem}

\begin{document}

\title{Information and Sufficiency on the Stock Market}

\author{\IEEEauthorblockN{Peter~Harremoës}\IEEEauthorblockA{Copenhagen Business College\\
Copenhagen\\
Denmark\\
Email: harremoes@ieee.org}}
\maketitle
\begin{abstract}
It is well-known that there are a number of relations between theoretical
finance theory and information theory. Some of these relations are
exact and some are approximate. In this paper we will explore some
of these relations and determine under which conditions the relations
are exact. It turns out that portfolio theory always leads to Bregman
divergences. The Bregman divergence is only proportional to information
divergence in situations that are essentially equal to the type of
gambling studied by Kelly. This can be related an abstract sufficiency
condition.
\end{abstract}

\section{INTRODUCTION}

The relation between gambling and information theory has been known
since Kelly \cite{Kelly1956}. Later Kelly's theory has been extended
to trading of assets, but the link to information theory is weaker
than in the case of gambling \cite{Cover1991}. In both gambling theory
and more general portfolio theory logarithmic terms appear because
we are interested in the exponential growth rate. In this paper we
shall demonstrate that portfolio theory consist of two parts. The
general part is related to Bregman divergences and this part is shared
with a number of other convex optimization problems. If a sufficiency
condition is imposed on the general theory we arrive at a theory where
the Bregman divergence reduces to information divergence. The sufficiency
is essentially equal to Kelly's theory of gambling.

The general theory of convex optimization and Bregman divergences
has a number of important applications. In each of the applications
we get a strong link to information theory if a sufficiency condition
is imposed. Therefore sufficiency conditions will lead to strong relations
between the different applications. 

In information theory an important goal is to compress. As long as
we restrict to uniquely decodable codes we get a Bregman divergence.
The sufficiency condition corresponds to allowing codewords real valued
length which is relevant when we allow block codes with no upper limit
on the block length. This leads to the wide spread use of information
divergence in information theory. The link between information divergence
and the notion of sufficency was emphazied already by Kullback and
Leibler in 1951 in the paper entitled ``Information and Sufficiency''
\cite{Kullback1951}. 

In statistics the idea of scoring rules has its roots in the 1920's
in the Dutch book theorem by Ramsay and de Finetti. McCathy \cite{McCarthy1956}
studied scoring rules in a more systematic way and Dawid, Lauritzen
and Parry \cite{Dawid2012} have recently extended the notion of proper
local scoring rules. Proper scoring rules leads to Bregman divergences
and sufficiency lead to local proper scoring rules. The basic result
is that any strictly local proper scoring rule is proportional to
logaritmic score. The link between information theory and statistics
is now very well established \cite{Csiszar2004}.

Convex optimization also appear in thermodynamics and statistical
mechanics where the goal is to extract as much energy as possible
from some physical system. The notion of entropy obviously play an
important role in both theories, but the best interpretation has been
debated ever since Shannon decided to call his quantity entropy. Since
all these theories are related we also get a link between finance
theory and physics so there is a whole topic called econophysics where
ideas from physics are applied to economic systems. We hope that the
present paper will help to understand to what extend quantities in
finance are really proportional to quantities in information theory,
statistics, or physics.

The general idea of using Bregman divergences for convex optimization
was presented in \cite{Harremoes2015a}. In the present paper we will
develop the theory further. Therefore there will be some overlap between
then the presentation in \cite{Harremoes2015a} and the present paper.
The second goal of this paper is apply the general theory to portfolio
theory.

\section{OPTIMIZATION}

Assume that our knowledge of a system can be represented by an element
in a convex set $S$ that we will call the \emph{state space}. The
simplest case of a state space is the simplex of probability measures
on a set. In quantum information theory the state space is the set
of density matrices on a Hilbert space. For states $s_{0}$ and $s_{1}$
and $t\in\left[0,1\right]$ the convex combination $\left(1-t\right)\cdot s_{0}+t\cdot s_{1}$
is identified with the mixed state where $s_{0}$ is taken with probability
$1-t$ and the state $s_{1}$ is taken with probability $t.$ The
\emph{pure states} are the extreme points of the state space. For
simplicity we will assume that the state space is a finite dimensional
convex compact set.

Let $\mathcal{A}$ denote a subset of the feasible measurements such
that $a\in A$ maps $S$ into a distribution on the real numbers i.e.
a random variable. The elements of $\mathcal{A}$ may represent \emph{actions}
(decisions) that lead to a payoff like the score of a statistical
decision, the energy extracted by a certain interaction with the system,
(minus) the length of a codeword of the next encoded input letter
using a specific code book, or the revenue of using a certain portfolio.
If the action $a$ is applied to the state $s$ then we get a random
variable $a\left(s\right)$ that we will allow to take values in $\mathbb{R}\cup\left\{ -\infty\right\} $.
For each $s\in\mathcal{S}$ we define $F\left(s\right)=\sup_{a\in\mathcal{A}}E\left[a\left(s\right)\right]$.
Without loss of generality we may assume that the set of actions $\mathcal{A}$
is closed so that we may assume that there exists $a\in\mathcal{A}$
such that $F\left(s\right)=E\left[a\left(s\right)\right]$ and in
this case we say that $a$ is optimal for $s.$ We note that $F$
is convex but $F$ need not be strictly convex. 
\begin{defn}
If $F\left(s\right)$ is finite \emph{the regret} of the action $a$
is defined by 
\begin{equation}
D_{F}\left(s,a\right)=F\left(s\right)-E\left[a\left(s\right)\right]
\end{equation}
\end{defn}
\begin{prop}
The regret $D_{F}$ has the following properties:\end{prop}
\begin{itemize}
\item $D_{F}\left(s,a\right)\geq0$ with equality if $a$ is optimal for
$s$.
\item If $\hat{a}$ is optimal for the state $\hat{s}=\sum t_{i}\cdot s_{i}$
where $\left(t_{1},t_{2},\dots,t_{\ell}\right)$ is a probability
vector then 
\[
\sum t_{i}\cdot D_{F}\left(s_{i},a\right)=\sum t_{i}\cdot D_{F}\left(s_{i},\hat{a}\right)+D_{F}\left(\hat{s},a\right).
\]
 
\item $\sum t_{i}\cdot D_{F}\left(s_{i},a\right)$ is minimal if $a$is
optimal for $\sum t_{i}\cdot s_{i}$.
\end{itemize}
If the state is $s_{1}$ but one acts as if the state were $s_{2}$
one suffers a regret that equals the difference between what one achieves
and what could have been achieved. 
\begin{defn}
If $F\left(s_{1}\right)$ is finite \emph{the regret} is defined by
\begin{equation}
D_{F}\left(s_{1},s_{2}\right)=\inf_{a}D_{F}\left(s,a_{n}\right)
\end{equation}
where the infimum is taken over actions $a$ that are optimal for
$s_{2}.$ 
\end{defn}
If there exists a unique action $a$ such that $F\left(s\right)=E\left[a\left(s\right)\right]$
then $F$ is differentiable which implies that the regret can be written
as a \emph{Bregman divergence} in the following form
\begin{align}
D_{F}\left(s_{1},s_{2}\right) & =F\left(s_{1}\right)-\left(F\left(s_{2}\right)+\left\langle s_{1}-s_{2},\nabla F\left(s_{2}\right)\right\rangle \right).
\end{align}
In the context of forecasting and statistical scoring rules the use
of Bregman divergences dates back to \cite{Hendrickson1971}. 

Bregman divergences satisfy the \emph{Bregman identity} 
\[
\sum t_{i}\cdot D_{F}\left(s_{i},\tilde{s}\right)=\sum t_{i}\cdot D_{F}\left(s_{i},\hat{s}\right)+D_{F}\left(\hat{s},\tilde{s}\right)
\]
but if $F$ is not differentiable this identity can be violated. If
the state $s_{2}$ has the unique optimal action $a_{2}$ then 
\begin{equation}
F\left(s_{1}\right)=D_{F}\left(s_{1},s_{2}\right)+E\left[a_{2}\left(s_{1}\right)\right]
\end{equation}
 so the function $F$ can be reconstructed from $D_{F}$ except for
an affine function of $s_{1}.$ Similarly the divergence $D_{F}$
is uniquely determined by the function $F.$

Consider the case where the state is not know exactly but we know
that $s\in\mathcal{S}$ for some set of states. The minimax regret
of the set $S$ is defined as 
\[
C_{F}=\inf_{a}\sup_{i}D_{F}\left(s_{i},a\right).
\]
Using general minimax results we get

\[
C_{F}=\sup_{\vec{t}}\inf_{a}\sum_{i}t_{i}\cdot D_{F}\left(s_{i},a\right)
\]
where the supremum is taken over all probability vectors $\vec{t}$
supported on $\mathcal{S}$. This result can improved.
\begin{thm}
If $\left(t_{1},t_{2},\dots,t_{n}\right)$ is a probability vector
on the states $s_{1},s_{2},\dots,s_{n}$ with $\bar{s}=\sum t_{i}\cdot s_{i}$
and $a_{opt}$ is the optimal action for $\bar{s}$ then 
\[
C_{F}\geq\inf_{a}\sum t_{i}\cdot D_{F}\left(s_{i},a\right)+D_{F}\left(\bar{s},a_{opt}\right).
\]
If $a$ is an action and $s_{opt}$ is optimal then 
\[
\sup_{i}D_{F}\left(s_{i},a\right)\geq C_{F}+D_{F}\left(s_{opt},a\right).
\]

\end{thm}

\section{SUFFICIENCY}

Let $\left(s_{\theta}\right)_{\theta}$ denote a family of states
and let $\Phi$ denote an affine transformation $\mathcal{S}\to\mathcal{T}$
where $\mathcal{S}$ and $\mathcal{T}$ denote state spaces. Then
$\Phi$ is said to be \emph{sufficient} for $\left(s_{\theta}\right)_{\theta}$
if there exists an affine transformation $\Psi:\mathcal{T}\to\mathcal{S}$
such that $\Psi\left(\Phi\left(s_{\theta}\right)\right)=s_{\theta}.$

We define a transformation $\Phi$ to be an \emph{isomixture }if $\Phi$
has the form $\Phi=\sum_{i=1}^{k}p_{i}\cdot\Phi_{i}$ where $\left(p_{1},p_{2},\cdots,p_{k}\right)$
is a probability vector and $\Phi_{i}$ is a isometry, i.e. a bijective
transformation of the state into itself. We say that the regret $D_{F}$
on the state space $S$ satisfies the \emph{iso-sufficiency property}
if 
\begin{equation}
D_{F}\left(\Phi\left(s_{1}\right),\Phi\left(s_{2}\right)\right)=D_{F}\left(s_{1},s_{2}\right)\label{eq:suff}
\end{equation}
for any isomixture $\mathcal{S}\to\mathcal{S}$ that is sufficient
for $\left(s_{1},s_{2}\right).$ The notion of sufficiency as a property
of divergences was introduced in \cite{Harremoes2007a}. The crucial
idea of restricting the attention to transformations of the state
space into itself was introduced in \cite{Jiao2014}.

The center of a convex set $S$ is the set of point in $S$ that are
invariant under isometries of $S.$ Note that the center is convex
and non-empty \cite{Lim1981}. If the center of the state space is
not a point there are many Bregman divergences that satisfy the sufficiency
condition.
\begin{prop}
Let $G$ denote the set of isometries of a state space $S$ and let
$\mu$ denote the Haar probability measure on $G.$ Let $\Phi$ denote
the projection $s\to\int g\left(s\right)\,\mathrm{d}\mu.$ Let $F$
denote a concave function on the center of $S.$ Then $D_{F}\left(\Phi\left(s_{1}\right),\Phi\left(s_{2}\right)\right)$
defines a Bregman divergence on $S$ that satisfies the iso-sufficiency
condition. 
\end{prop}

\begin{prop}
\label{thm:symm}Assume that $S$ is a state space. If the divergence
$D_{F}$ satisfies the iso-sufficiency property then there exists
a $\tilde{F}$ such that 
\[
D_{\tilde{F}}\left(s_{1},s_{2}\right)=D_{F}\left(s_{1},s_{2}\right)
\]
and $\tilde{F}\left(\Phi\left(s\right)\right)=\tilde{F}\left(s\right)$
.
\end{prop}
If the state space is a one dimensional simplex then the only sufficient
transformation is the reflection and the above condition on $F$ is
sufficient to conclude that Equation \ref{eq:suff} holds.
\begin{prop}
If the state space has the shape of a ball then any function $F$
on the ball that is concave and invariant under rotations satisfies
the iso-sufficiency condition.\end{prop}
\begin{IEEEproof}
Assume that the isomixture $\Phi$ is sufficient for $\left\{ s_{0},s_{1}\right\} .$
Then $\Phi$ is also sufficient for any affine conbination of $s_{0}$
and $s_{1}$. In particular we may replace $s_{0}$ and $s_{1}$ by
affine combinations for the form $s_{t_{i}}=\left(1-t_{i}\right)\cdot s_{0}+t_{i}\cdot s_{1}$
that are extreme points in $\mathcal{S}.$ Since $\Phi$ is assumed
to be sufficient it maps $s_{t_{i}}$ into an extreme points. Hence
$\Phi$ acts as a rotation on the intersection of the state space
and the affine span of $s_{1},s_{2}$ and $U$. Since $F$ is invariant
under rotations the divergence $D_{F}$ is also invariant under rotations
implying that $D_{F}\left(\Phi\left(s_{1}\right),\Phi\left(s_{2}\right)\right)$=$D_{F}\left(s_{1},s_{2}\right).$ 
\end{IEEEproof}
The simplest case of a ball is an interval, which corresponds to the
probability measures on a binary alphabet. This special case was discussed
in \cite{Jiao2014}. The balls in dimensions 2, 3, and 5 correspond
to density matrices of a 2 dimensional Hilbert space over the real
numbers, over the complex numbers, and over the quarternions. 

We say that the states $s_{0}$ and $s_{1}$ are \emph{orthogonal}
and write $s_{1}\bot s_{2}$ if there exists an affine function $\phi:S\to\left[0,1\right]$
such $\phi\left(s_{0}\right)=0$ and $\phi\left(s_{1}\right)=1.$
The following theorem can be proved by the same technique as \cite[Thm. 4]{Harremoes2015a}
except that we will make sufficient projections by taking the mean
actions of a groups equipped with the Haar probability measure.
\begin{thm}
\label{thm:Generelt}Assume that the state space $S$ satisfies the
following properties:

1. For and two pure states $s_{1}$ and $s_{2}$ there exists an isometry
of $S$ such that $\Phi\left(s_{1}\right)=s_{2}.$

2. For any three pure states $s_{1},s_{2},$ and $s_{3}$such that
$s_{1}\bot s_{3}$ and $s_{2}\bot s_{3}$ there exists an isometry
of $S$ such that $\Phi\left(s_{1}\right)=s_{2}$ and $\Phi\left(s_{3}\right)=s_{3}.$

3. The state space has at least three orthogonal pure states.

4. Any state can be written as a mixture of orthogonal pure states.

If the regret $D_{F}$ satisfies the iso-sufficiency property given
by Equation \ref{eq:suff}, then $D_{F}$ is uniquely determined except
for a multiplicative factor.\end{thm}
\begin{rem}
Condition 4 seems to be redundant, but we have not been able to prove
this.
\end{rem}
When the state space is a simplex the uniquely determined divergence
is information divergence and when the state space is density matrices
on a complex Hilbert space we get quantum relative entropy.
\begin{lem}
\label{lem:ortho}Assmue that the state space satisfies the conditions
in Theorem \ref{thm:Generelt}. If $s_{0}\bot s_{1}$ then any optimal
action $a$ for $s_{1}$ satisfies $E\left[a\left(s_{0}\right)\right]=-\infty.$\end{lem}
\begin{IEEEproof}
Since $s_{0}$ and $s_{1}$ are orthogonal and the conditions in the
previous theorem is fulfilled the we have that the regret restricted
to the line segment $\left\{ t\in\left[0,1\right]\mid\left(1-t\right)s_{0}+ts_{1}\right\} $
is proportional to information divergence, but information divergence
equals $\infty$ for orthogonal distributions so $D_{F}\left(s_{0},s_{1}\right)=\infty.$
Hence $\inf_{a}\left(F\left(s_{0}\right)-E\left[a\left(s_{0}\right)\right]\right)=\infty$
where the infoimum is taken over actions that are optimal for $s_{1}$.
Therefore $E\left[a\left(s_{0}\right)\right]=-\infty$ for any action
$a$ that is optimal for $s_{1}.$
\end{IEEEproof}

\section{Portfolio theory}

Let $X_{1},X_{2},\dots,X_{k}$ denote \emph{price relatives} for a
list of $k$ assets. For instance $X_{5}=1.04$ means that asset no.
5 increases its value by 4 \%. 
\begin{example}
A special asset is the \emph{safe asset} where the price relative
is 1 for any feasible price relative vector. Investing in this asset
corresponds to place the money at a safe place with interest rate
equal to 0 \% .
\end{example}
A \emph{portfolio} is an asset given by a probability vector $\vec{b}=\left(b_{1},b_{2},\dots,b_{k}\right)$
where for instance $b_{5}=0.3$ means that 30 \% of the money is invested
in asset no. 5. The total price relative is $X_{1}\cdot b_{1}+X_{2}\cdot b_{2}+\dots+X_{k}\cdot b_{k}=\left\langle \vec{X},\vec{b}\right\rangle .$
If an asset has the property that the price relative is only positive
for one of the feasible price relative vectors, then we may call it
a \emph{gambling asset}. For any set of \emph{possible assets} we
may extend the set of assets by a number of \emph{ideal gambling assets}
so that any of the possible assets can be written as a portfolio of
the ideal gambling assets. This can be done without changing the set
of feasible price relative vectors. Therefore the set of possible
portfolios may be considered as a convex subset of a set of portfolios
of some ideal gambling assets. 

We now consider a situation where the assets are traded once every
day. For a sequence of price relative vectors $\vec{X}_{1},\vec{X_{2}},\dots\vec{X}_{n}$
and \emph{a constant re-balancing portfolio} $\vec{b}$ the wealth
after $n$ days is
\begin{eqnarray}
S_{n} & = & \prod_{i=1}^{n}\left\langle \vec{X}_{i},\vec{b}\right\rangle \\
 & = & \exp\left(\sum_{i=1}^{n}\log\left(\left\langle \vec{X}_{i},\vec{b}\right\rangle \right)\right)\\
 & = & \exp\left(n\cdot E\left[\log\left\langle \vec{X},\vec{b}\right\rangle \right]\right)
\end{eqnarray}
where the expectation is taken with respect to the empirical distribution
of the price relative vectors. Here $E\left[\log\left\langle \vec{X},\vec{b}\right\rangle \right]$
is proportional to the \emph{doubling rate} and is denoted $W\left(\vec{b},P\right)$
where $P$ indicates the probability distribution of $\vec{X}$. Our
goal is to maximize $W\left(\vec{b},P\right)$ by choosing an appropriate
portfolio $\vec{b}.$ In \cite{Cover1991} and \cite{Harremoes2015a}
it was tacitly assumed that a unique optimal portfolio exists but
this is not always the case. Here we will not assume uniqqueness. 
\begin{defn}
Let $\vec{b}_{1}$ and $\vec{b}_{2}$ denote two portfolios. We say
that $\vec{b}_{1}$ \emph{dominates} $\vec{b}_{2}$ if $\left\langle \vec{X}_{j},\vec{b}_{1}\right\rangle \geq\left\langle \vec{X}_{j},\vec{b}_{2}\right\rangle $
for any $j=1,2,\dots,n.$ We say that $\vec{b}_{1}$ \emph{strictly
dominates} $\vec{b}_{2}$ if $\left\langle \vec{X}_{j},\vec{b}_{1}\right\rangle >\left\langle \vec{X}_{j},\vec{b}_{2}\right\rangle $
for any $j=1,2,\dots,n.$
\end{defn}
For a vector $\vec{v}=\left(v_{1},v_{2},\dots,v_{k}\right)\in\mathbb{R}^{k}$
the support $supp\left(\vec{v}\right)$ is the set of indices $i$
such that $v_{i}>0.$ We note that if $\vec{b}_{1}$ strictly dominates
$\vec{b}_{2}$ if and only if there exists an $i\in supp\left(\vec{b}_{2}\right)$
such that $\vec{b}_{1}$ strictly dominates $\vec{e}_{i}$ where $\vec{e}_{i}$
denotes the $i$'th basis vector. The consequence is that we may remove
assets number $i$ if $\vec{e}_{i}$ is strictly dominated because
one will never put any money on that particular asset. Similarly,
$\vec{b}_{1}$ dominates $\vec{b}_{2}$ if and only if there exists
an $i\in supp\left(\vec{b}_{2}\right)$ such that $\vec{b}_{1}$ dominates
$\vec{e}_{i}$. We do not decrease the maximal doubling rate by removing
assets that are dominated, but sometimes assets that are dominated
but not strictly dominated may lead to non-uniqueness of the optimal
portfolio. 
\begin{defn}
A set $A$ of assets is said to dominate the set of assets $B$ if
any asset in $B$ is dominated by a by a portfolio of assets in $A.$\end{defn}
\begin{prop}
If $\vec{b}_{0}$ is optimal for the distribution $\delta_{\vec{v}}$
then the support of $\vec{b}$ is a subset of the support of $\vec{v}.$ \end{prop}
\begin{IEEEproof}
If $P=\delta_{\vec{v}}$ then $E\left[\log\left\langle \vec{X},\vec{b}\right\rangle \right]=\log\left\langle \vec{v},\vec{b}\right\rangle .$
The portfolio $\vec{b}$ is a probability distribution over stocks
so if we let $\vec{b}*$ denote the conditional distribution of $\vec{b}$
on the support of $\vec{v}.$ Then 
\[
\log\left\langle \vec{v},\vec{b}_{0}\right\rangle \leq\log\left\langle \vec{v},\vec{b}*\right\rangle 
\]
 with equality if and only if the support of $\vec{b}$ is a subset
of the support of $\vec{v}.$ Therefore $\vec{b}=\vec{b}_{0}$ implies
that the support of $\vec{b}$ is a subset of the support of $\vec{v}.$
\end{IEEEproof}
Let $\vec{b}_{P}$ denote a portfolio that is optimal for $P$. The
regret of choosing a portfolio according to $Q$ when the distribution
is $P$ is given by the Bregman divergence 
\[
W\left(\vec{b}_{P},P\right)-W\left(\vec{b}_{Q},P\right).
\]
If $\vec{b}_{Q}$ is not uniquely determined we take a minimum over
all $\vec{b}$ that are optimal for $Q.$
\begin{example}
If the assets are orthogonal gambling assets we get the type of gambling
described by Kelly. There will be one-to-one correspondence between
price relative vectors and assets. For a probability disttribution
$P$ over price relative vectors the optimal portfolio $\vec{b}_{P}$
is a vector with the same coordinates as the probability vector $P.$
We have 
\begin{equation}
W\left(\vec{b}_{P},P\right)-W\left(\vec{b}_{Q},P\right)=D\left(P\Vert Q\right)\label{eq:perfekt}
\end{equation}
so the sufficiency condition is fulfilled in gambling.
\end{example}
If a set of possible assets it embedded as a subset $C$ in a set
of ideal gambling assets then $C$ may be identified with a convex
set of probability distributions. Now maximizing $W\left(\vec{b},P\right)$
over possible portfolios $\vec{b}$ is the same as minimizing the
regret given by (\ref{eq:perfekt}) over $Q\in C$ in the set of portfolios
over ideal gambling assets. Therefore $\vec{b}_{Q}$ may be identified
with a reversed information projection of $Q$ on $C.$

As proved in \cite{Cover1991} the regret satisfies
\begin{equation}
W\left(\vec{b}_{P},P\right)-W\left(\vec{b}_{Q},P\right)\leq D\left(\left.P\right\Vert Q\right).\label{eq:Cover}
\end{equation}
In the set of portfolios over ideal assets there is a on-to-one correspondence
between mixed states and portfolios. Therefore maximizing $W\left(\vec{b},P\right)$
over $\vec{b}$ in the original set of portfolios corresponds to minimizing
the regret $W\left(\vec{b}_{P},P\right)-W\left(\vec{b}_{Q},P\right)$
over $Q$ which again corresponds to minimizing $D\left(P\Vert Q\right)$
under the condition that $\vec{b}_{Q}\in C$ in a set of portfolios
on orthogonal gambling assets. The inequality (\ref{eq:Cover}) therefore
states that information divergence decreases when probability measures
are projected (reverse information projection) into a convex set.
Here we should note that information divergence is convex but not
strictly convex in the second argument. Therefore the reversed information
may be non-unique.

\section{Sufficient portfolios}
\begin{lem}
\label{lem:todim}Assume that there are only two price relative vectors
and that the set of assets is minimal dominating. If the Bregman divergence
\begin{equation}
W\left(\vec{b}_{P},P\right)-W\left(\vec{b}_{Q},P\right)
\end{equation}
 is proportional to information divergence $D\left(\left.P\right\Vert Q\right)$
then there are only two gambling assets.\end{lem}
\begin{IEEEproof}
Let 
\begin{eqnarray*}
\vec{X} & = & \left(X_{1},X_{2},\dots,X_{k}\right)\\
\vec{Y} & = & \left(\,Y_{1},\,Y_{2},\,\dots,\,Y_{k}\right)
\end{eqnarray*}
denote the two price relative vectors. If $P=\left(s,t\right)$ then
the vector $\vec{b}=\left(b_{1},b_{2},\dots,b_{n}\right)$ is log-optimal
if and only if 
\begin{multline*}
s\frac{X_{i}}{b_{1}X_{1}+\dots+b_{k}X_{k}}+t\frac{Y_{i}}{b_{1}Y_{1}+\dots+b_{k}Y_{k}}\leq1
\end{multline*}
for all $i\in\left\{ 1,2,\dots,k\right\} $ with equality if $b_{i}>0.$
Since we have assumed that none of the assets are dominated by other
portfolios only two of these inequalities can hold with equality.
Therefore we may assume that only $b_{1}$ and $b_{2}$ are positive.
Hence we may assume that there are only two assets.

Let $\delta_{1}$ denote the measure concentrated on $\vec{X}$ and
let $\delta_{2}$ denote the measure concentrated on $\vec{Y}.$ Since
the measures $\delta_{1}$ and $\delta_{2}$ are orthogonal Lemma
\ref{lem:ortho} we have that $W\left(\vec{b}_{\delta_{j}},\delta_{i}\right)=-\infty.$
Now 
\begin{eqnarray*}
W\left(\vec{b}_{\delta_{j}},\delta_{i}\right) & = & E_{\delta_{i}}\left[\log\left\langle \vec{X},\vec{b}_{\delta_{j}}\right\rangle \right]\\
 & = & \log\left\langle \vec{X}_{i},\vec{b}_{\delta_{j}}\right\rangle 
\end{eqnarray*}
so that $\left\langle \vec{X}_{i},\vec{b}_{\delta_{j}}\right\rangle =0$.
Since the support of $\vec{b}_{\delta_{i}}$ is a subset of the support
of $\vec{X}_{i}$ we have that $\vec{b}_{\delta_{i}}\bot\vec{b}_{\delta_{j}}.$
Therefore $\vec{b}_{\delta_{1}}$ and $\vec{b}_{\delta_{2}}$ must
be proportional to the basis vectors. Since $\vec{b}_{\delta_{1}}$
and $\vec{b}_{\delta_{2}}$ are vectors in a $2$-dimensional space
and their coordinates are non-negative we have that $\vec{b}_{\delta_{i}}$
must proportional to a basis vector. Since $\left\langle \vec{X}_{i},\vec{b}_{\delta_{j}}\right\rangle =0$
for $i\neq j$ we have that $\vec{X}_{i}$ is parallel with $\vec{b}_{\delta_{i}}.$\end{IEEEproof}
\begin{thm}
\label{thm:proper}Assume that none of the assets are dominated by
a portfolio of the other assets. If the Bregman divergence 
\begin{equation}
W\left(\vec{b}_{P},P\right)-W\left(\vec{b}_{Q},P\right)
\end{equation}
 is proportional to information divergence $D\left(\left.P\right\Vert Q\right)$
the measures $P$ and $Q$ are supported by $k$ distinct price relative
vectors of the form $\left(o_{1},0,0,\dots0\right)$, $\left(0,o_{2},0,\dots0\right),$
until $\left(0,0,\dots o_{k}\right).$ \end{thm}
\begin{IEEEproof}
Assume that there exists a constant $c>0$ such that 
\begin{equation}
W\left(\vec{b}_{P},P\right)-W\left(\vec{b}_{Q},P\right)=c\cdot D\left(\left.P\right\Vert Q\right).\label{eq:propertional}
\end{equation}
 If $\vec{b}_{P}=\vec{b}_{Q}$ then 
\[
W\left(\vec{b}_{P},P\right)-W\left(\vec{b}_{Q},P\right)=0
\]
 and $D\left(\left.P\right\Vert Q\right)=0$ and $P=Q.$ Therefore
the mapping $P\to\vec{b}_{P}$ is injective. The vectors $\vec{b}_{P}$
form a simplex with $k$ extreme points. Therefore the simplex of
probability measures $P$ has at most $k$ extreme points, so $P$
is supported on at most $k$ distinct vectors that we will denote
$\vec{X}_{1},\vec{X}_{2},\dots,\vec{X}_{k}$. 

Assume that $\vec{X}$ and $\vec{Y}$ are two vectors of price relatives.
Then Equation \ref{eq:propertional} holds for probability vectors
restricted to the set $\left\{ \vec{X},\vec{Y}\right\} .$ From Lemma
\ref{lem:todim} it follows that $\vec{X}$ and $\vec{Y}$ are orthogonal.
Therefore all the price relative vectors are orthogonal, and have
disjoint supports. Since the price relative vectors have disjoint
support, an asset can only have a positive price relative for one
of the price relative vectors. Therefore each price relative vector
has one asset that dominates any other asset in the support of the
price relative vector. Since we have assumed none of the assets are
dominated each price relative vector is supported on a single asset.

If the price relative vectors are as in Theorem \ref{thm:proper}
we are in the situation of gambling introduced by Kelly \cite{Kelly1956}. \end{IEEEproof}
\begin{cor}
Assume that the Bregman divergence 
\begin{equation}
W\left(\vec{b}_{P},P\right)-W\left(\vec{b}_{Q},P\right)\label{eq:Bregman}
\end{equation}
satisfies the sufficiency condition for probability measures $P$
and $Q$ supported on $k\geq3$ price relative vectors. Then the set
of possible assets contain $k$ gambling assets and any other asset
is dominated by a portfolio on the gambling assets.\end{cor}
\begin{example}
If the Breman divergence satisfies the sufficiency condition and one
of the assets is the safe asset then there exists a portfolio $\vec{b}$
such that $b_{i}\cdot o_{i}\geq1$ for all $i.$ Equivalently $b_{i}\geq o_{i}^{-1}$
which is possible if and only if $\sum o_{i}^{-1}\leq1.$ One say
that the gamble is \emph{fair} if $\sum o_{i}^{-1}=1$. If the gamble
is \emph{superfair}, i.e. $\sum o_{i}^{-1}<1$, then the portfolio
$b_{i}=o_{i}^{-1}/\sum o_{i}^{-1}$ gives a price relative equal to
$\left(\sum o_{i}^{-1}\right)^{-1}>1$ independetly of what happens,
which is a \emph{Dutch book}. \end{example}
\begin{cor}
Assume that there are at least three distinct price relative vectors.
The Bregman divergence (\ref{eq:Bregman}) satisfies the sufficiency
doncition if and only if $W\left(\vec{b}_{P},P\right)-W\left(\vec{b}_{Q},P\right)=0$
implies $P=Q.$\end{cor}
\begin{IEEEproof}
If Equation \ref{eq:perfekt} does not hold then we do not have sufficiency
so the set of possible portfolios can be identified with a convex
and proper subset of the set of all portfolios on a set of gambling
assets. Then we just have to find to distributions $P$ and $Q$ that
have the same reversed information projection into the set of possible
portfolios.
\end{IEEEproof}

\section{CONCLUSION}

The link between portfolio theory and information theory works on
two levels. Parts of the theory can be stated and proved on the level
of convex optimization, where Bregman divergences and related concepts
play a central role. If we further impose a sufficiency condition
we have, essentially, to restrict our attention to gambling as described
by Kelly. Adding certain assets that are dominated does not make any
significant changes to the theory. In the case of gambling the correspondence
between portfolio theory and information theory becomes perfect. Therefore
the link between general portfolio theory and information theory is
convayed by gambling theory. 

Information divergence was introduced by Kullback and Leibler in the
paper entitled ``On Information and Sufficiency''. In the present
paper we have made the notion of sufficiency more explicit for portfolio
theory. The introduction of ideal gambling assets paralellels the
use of microscopic states as opposed to macroscopic states in physics.
For microscopic states we have reversibility and conservation of energy.
Similarly, gambling corresponds to two-person zero sum games where
money is the conserved quantity. As we have seen these correspondencies
are consequences of the sufficiency condition.

\section*{Acknowledgement}

Tha author want to thank Prasad Santhanam for inviting me to Electical
Engineering Department, University of Hawai'i, where this paper was
written. 

\bibliographystyle{IEEEtran}
\bibliography{C:/Users/Peth/Documents/bibtex/database1}

\end{document}